\documentclass[a4paper,11pt]{article}
\usepackage{pos}

\usepackage{slashed}
\usepackage{upgreek}
\usepackage[normalem]{ulem}
\global\long\def\dd{\mathrm{d}}
\global\long\def\Vol{\mathrm{Vol}}

\title{Deforming ${\rm AdS}_3\times S^3\times T^4$ in Type IIB Supergravity}

\author[b,c]{Stefano Maurelli}
\author[a,b]{Ruggero Noris}
\author[d]{Marcelo Oyarzo}
\author*[b,c]{Mario Trigiante}
\affiliation[a]{CEICO, Institute of Physics of the Czech Academy of Sciences,
Na Slovance 2, 182 21 Prague 8, Czech Republic}
\affiliation[b]{Department of Applied Science and Technology, Politecnico di Torino, Corso Duca degli Abruzzi, 24, 10129 Torino, Italy}
\affiliation[c]{INFN, Sezione di Torino, Via P. Giuria 1, 10125 Torino, Italy}
\affiliation[d]{
Instituto Galego de F\'isica de Altas Enerx\'ias (IGFAE), Universidade de Santiago de Compostela, E-15782 Santiago de Compostela, Spain.}
\emailAdd{ruggeronoris28@gmail.com}
\emailAdd{stefano.maurelli@polito.it}
\emailAdd{moyarzoca1@gmail.com}
\emailAdd{mario.trigiante@polito.it}
\abstract{We discuss some new results on the construction of supersymmetric solutions of Type IIB supergravity of the form ${\rm WAdS}_3\times{\rm WS}^3\times T^4$, ${\rm WAdS}_3$ and ${\rm WS}^3$ denoting \emph{warped} anti-de Sitter spacetime and sphere, respectively. The distinctive feature of these backgrounds is that, in spite of them being supersymmetric,  the warpings of the two factors are described by independent parameters. We illustrate how some of these geometries, characterised by a lightlike warping of the anti-de Sitter factor, arise in the near-horizon limit of a regular,  asymptotically locally flat configuration of D-branes and fluxes. Central to the construction of the latter solutions is the use of two independent TsT transformations. We also give a new class of supersymmetric solutions  of the general form ${\rm WAdS}_3\times{\rm WS}^3\times T^4$, which has not been published yet. They feature warpings of the anti-de Sitter factor of the lightlike, spacelike and timelike types. We discuss their properties.}

\FullConference{Proceedings of the Corfu Summer Institute 2025 "School and Workshops on Elementary Particle Physics and Gravity" (CORFU2025)\\
27 April - 28 September, 2025\\
Corfu, Greece\\}

\begin{document}
\maketitle
\section{Introduction}
The conjectured holographic duality between superstring/M-theories on an asymptotically anti-de Sitter (AAdS) background and a quantum CFT on the boundary of the AAdS geometry \cite{Maldacena:1997re,Gubser:1998bc,Witten:1998qj} has been one of the most fruitful ideas of the last decades. It has allowed to gain important insights into both sides of the duality. In particular, taking the supergravity limit of superstring theory, it led to a deeper understanding of the strong-coupling regime of the dual, non-gravitational theory.
\par
Type IIB superstring solutions of the form ${\rm AdS}_3\times S^3\times M_4$, $M_4=K_3$ or $T^4$, have represented a privileged testing ground for the AdS/CFT duality for several reasons. These backgrounds describe the near-horizon geometry of D1-D5 (or F1-NS5) systems, and on them the first microscopic entropy countings of a black hole in superstring theory were achieved, yielding the expected Bekenstein-Hawking entropy formula \cite{Strominger:1996sh,Strominger:1997eq}. Superstring theory on such backgrounds is dual to a 1+1 SCFT  on the boundary of ${\rm AdS}_3$, which is conjectured to be a (deformation of) the orbifold model ${\rm Sym}^N(M_4)$. The infinite conformal symmetry of the boundary theory was first derived from the asymptotic symmetries on ${\rm AdS}_3$ in \cite{Brown:1986nw}.
This holographic duality has been extensively studied \cite{Antoniadis:1989mn,Maldacena:1998bw,Deger:1998nm,deBoer:1998kjm} (for a recent review, see \cite{Kovensky:2026usc}), even beyond the supergravity limit, since ${\rm AdS}_3\times S^3\times M_4$ backgrounds can also be realised with NS-NS fields only and string theory on them is formulated as an ${\rm SU}(2)\times {\rm SU}(2)$-WZW model \cite{Antoniadis:1989mn,Maldacena:2000hw,Maldacena:2000kv,Maldacena:2001km, Eberhardt:2019ywk}.\par
A significant effort has been made towards extending this correspondence to variants of the ${\rm AdS}_3\times S^3\times M_4$ backgrounds, including three-dimensional spacetime geometries in which even the local, asymptotic anti-de Sitter symmetry is broken. A natural question to ask is: among this kind of deformations of the ${\rm AdS}_3$ factor,  what are those which still retain a ``large amount'' of the original asymptotic isometries? The ``most symmetric'' deformations of ${\rm AdS}_3$ and $S^3$ are obtained by \emph{warping} these geometries, thus obtaining a \emph{warped}-${\rm AdS}_3$ (WAdS$_3$) and a \emph{warped}-${S}^3$ (WS$^3$) spaces, respectively. These are 1-parameter deformations of ${\rm AdS}_3$ and $S^3$,  which retain, in the former case,  an ${\rm SL}(2,\mathbb{R})\times {\rm U}(1)$ subgroup of the original  ${\rm SL}(2,\mathbb{R})\times {\rm SL}(2,\mathbb{R})$ isometry group and, in the latter case, an ${\rm U}(2)$ isometry group. The WAdS$_3$ isometry algebra $\mathfrak{sl}(2,\mathbb{R})\oplus \mathfrak{u}(1)$ is enhanced, asymptotically, to the direct sum of a Virasoro and a ${\rm U}(1)$-Kac-Moody algebras. These spacetimes (see \cite{Bieliavsky:2024vry} for a review of their geometry) and black holes asymptoting them, have been extensively studied in the literature in various contexts, such as Topologically Massive Gravity, Minimally Massive Gravity (see, for instance, \cite{Nutku:1993eb,Moussa:2003fc,Anninos:2008fx,Compere:2009zj,Chow:2009km,Nam:2018gju,Setare:2017xlu}) or (super-)gravity with torsion \cite{Andrianopoli:2023dfm, Andrianopoli:2024twc}.\par
Within superstring theory, backgrounds of the form ${\rm (W)AdS}_3\times {\rm (W)S}^3\times M_4$, in which either the anti-de Sitter or the sphere or both are warped, were proven to be exact string solutions, obtained from integrable marginal ``$J\bar{J}$'' deformations of the WZW model \cite{Israel:2004cd,Detournay:2005fz,Azeyanagi:2012zd}. \footnote{General deformations of $\mathrm{AdS}_3\times \mathrm S^3\times T^4$, not necessarily originating from TsT transformations, with special relation to the integrability of the underlying worldsheet model, have been extensively studied in the literature. See for instance \cite{Borsato:2018spz,Seibold:2019dvf,Hoare:2023zti,Hoare:2025rtl,Seibold:2025fnu,Borsato:2016ose,Itsios:2023kma}. For a recent review, see also \cite{Seibold:2024qkh}.} This is related to the fact that these backgrounds can be obtained from the undeformed one through a TsT transformation \cite{Lunin:2005jy}, i.e. a special ${\rm O}(d,d)$ symmetry which characterises a class of classical supergravity spacetimes featuring $d$ spatial translational symmetries (cyclic coordinates), and which extends the ${\rm O}(d,d;\,\mathbb{Z})$ symmetry of string theory. These transformations realise, at the level of the target space, the effect of marginal deformations of the worldsheet theory \cite{Orlando:2019rjg}.
In Type IIB supergravity, various (supersymmetric) ${\rm (W)AdS}_3\times {\rm (W)S}^3\times M_4$ backgrounds have been constructed, starting from the six-dimensional consistent truncation which describes the ten-dimensional theory compactified on $M_4$. In these cases, the warpings follow from the back-reaction on the geometry of appropriate fluxes \cite{El-Showk:2011euy,Hoare:2022asa,Eloy:2021fhc}. In particular, doubly warped ${\rm WAdS}_3\times {\rm WS}^3\times M_4$, 1-parameter solutions were constructed from a TsT transformation within a $D=6,\,\mathcal{N}=(2,0)$ supergravity which represents a consistent truncation of the Type IIB theory on these backgrounds \cite{El-Showk:2011euy,Hoare:2022asa}.
As far as 2-parameter solutions of this kind are concerned, non-supersymmetric examples were worked out in \cite{Orlando:2010ay}. \par
On this class of backgrounds, since the asymptotic conformal symmetry is broken, (super-) gravity is not expected to be dual to a CFT. The dual theory, referred to as \emph{Warped CFT}, has been characterised as the IR limit of a \emph{dipole-deformed field theory} and, as such, is non-local and lacks Lorentz invariance as well \cite{Bergman:2000cw,Dasgupta:2000ry,Bergman:2001rw,El-Showk:2011euy,Song:2011sr,Detournay:2012pc}. Aside from this conjectured duality, a precise holographic field-operator dictionary is still missing.\par
In \cite{Maurelli:2025iba}, a new class of doubly-warped 2-parameter solutions of Type IIB supergravity were constructed, on which the warpings of the anti-de Sitter part and of the sphere are independent.
This class includes doubly-warped supersymmetric backgrounds and was further extended in \cite{Maurelli:2025ueo}. Moreover, in the latter paper, we addressed the question of whether these new supersymmetric backgrounds could be obtained as a suitable near-horizon limit of an asymptotically flat, supersymmetric, regular brane system with fluxes. The latter interpolating configuration is constructed by applying two TsT transformations to the D1-D5 solution.\footnote{See \cite{Georgescu:2024iam} for a study of backgrounds obtained from a TsT transformation and dualities in the D1-D5 solution.} In what follows, we shall review this construction and give new, yet unpublished, ${\rm WAdS}_3\times {\rm WS}^3\times M_4$, 2-parameter, supersymmetric solutions. To our knowledge, the class of new solutions presented in \cite{Maurelli:2025iba} and \cite{Maurelli:2025ueo} contains the first supersymmetric backgrounds of the form  ${\rm WAdS}_3\times {\rm WS}^3\times M_4$, with independent warpings on the two factors. \par
We start by defining the main ingredients of our discussion, namely the spaces $ {\rm WAdS}_3,\,{\rm WS}^3$ and the TsT transformations. We then review the construction of the interpolating geometry, and discuss its properties. Finally we shall discuss the new solutions and end with some concluding remarks.

\section{Warping ${\rm AdS}_3$ and $S^3$}
In this section, we shall give a somewhat unified treatment of the warpings of these two spaces, referring to \cite{Maurelli:2025iba} for the relevant notations.
Both ${\rm AdS}_3$ and $S^3$ are symmetric Einstein spaces, with isometry groups ${\rm SL}(2,\,\mathbb{R})_L\times {\rm SL}(2,\,\mathbb{R})_R$ and ${\rm SU}(2)_L\times {\rm SU}(2)_R$, respectively. Let us generically denote by $K_{(L/R)x}{}^\upalpha$, $x=1,2,3$ and $\upalpha$ labelling the coordinates of any of the two spaces, the Killing vectors of ${\rm SL}(2,\,\mathbb{R})_L,\, {\rm SL}(2,\,\mathbb{R})_R$ or ${\rm SU}(2)_L,\,{\rm SU}(2)_R$, and let $g_{\upalpha\upbeta}$ generically denote the metric on such spaces. We can then define the following 1-forms ${\bf A}_{(L/R)\,x}\equiv K_{(L/R)x}^\upalpha\,g_{\upalpha\upbeta}\,\dd x^\upbeta$. These forms satisfy the following  Beltrami-like equations:
$$\star_3\, \dd {\bf A}_{(L)\,x}=\,m\, {\bf A}_{(L)\,x}\,,\,\,\,\star_3\, \dd {\bf A}_{(R)\,x}=\,-m\, {\bf A}_{(R)\,x}\,\,,\,\,\,\,m=\frac{2}{\ell}\,,$$
$\star_3$ denoting the Hodge dual on any of the two spaces and $\ell$ their radius. The sign on the right-hand side defines the`chirality' of the 1-forms. Warping the ${\rm AdS}_3$ or the  $S^3$ geometries amounts to choosing one particular 1-form ${\bf A}=A_{\upalpha}\,\dd x^\upalpha$ among the six ${\bf A}_{(L/R)\,x}$ and deforming the metric as follows:
$$g_{\upalpha\upbeta}\,\longrightarrow\,\,\,\,\,g_{\upalpha\upbeta}^{(W)}=g_{\upalpha\upbeta}+\varpi\,A_\upalpha\,A_\upbeta\,,$$
$\varpi$ being a constant parameter.
The isometry group breaks to the ${\rm SL}(2,\mathbb{R})\times {\rm U}(1)$ and ${\rm SU}(2)\times {\rm U}(1)$  subgroups of the isometry groups of ${\rm AdS}_3$ and $\mathrm{S}^3$, respectively, commuting with the Killing vector chosen for the deformation. The Ricci tensor of the deformed geometry has the characteristic form:
\begin{equation}
    \mathcal{R}_{\upalpha\upbeta}[g^{(W)}]=a(\varpi)\,g_{\upalpha\upbeta}^{(W)}+b(\varpi)\,A_\upalpha\,A_\upbeta\,.
\end{equation}
In the undeformed limit $\varpi\rightarrow 0$, $b(\varpi)\rightarrow 0$ and $a(\varpi)\rightarrow -2/\ell^2$.
Spaces of this kind are called \emph{K-contact}, \emph{$\eta$-Einstein} spaces. An equivalent way of performing this warping is to write the original metrics as Hopf fibrations and to rescale the metric along the fibre.\par While there is just one way of warping $S^3$, as $A_\upalpha$ on it can only be spacelike, there are three inequivalent ways of warping ${\rm AdS}_3$, depending on whether $A_\upalpha$ on ${\rm AdS}_3$ is timelike (\emph{timelike warped space}, denoted by WAdS$_3^-$), spacelike (\emph{spacelike warped space} WAdS$_3^+$) or null (\emph{lightlike warped space} WAdS$_3^0$). The null warped space WAdS$_3^0$ coincides with the three-dimensional Schroedinger space ${\rm Sch}_3^{(n=2)}$ with dynamical exponent $n=2$ \cite{Bobev:2011qx}.\footnote{See \cite{Son:2008ye} for an application fo this background to nonrelativistic holography, and \cite{Anninos:2010pm} for a study of the perturbative stability of black holes on these spaces within TMG.}
In this case, the absolute value of the warping parameter can be reabsorbed in a coordinate redefinition, so that $\varpi$ is an on/off parameter.\par
Since, with the exception of ${\rm WAdS}_3^0$, $\dd {\bf A}\wedge {\bf A}=\pm m\,\star_3{\bf A}\wedge {\bf A}\neq 0$, ${\bf A}$ defines a \emph{contact structure} on the corresponding three-dimensional space. Therefore, with a misuse of language, $ {\bf A}$ will always be referred to as a \emph{contact 1-form}, although, strictly speaking, on ${\rm WAdS}_3^0$ it is not.
\section{TsT Transformations}
A subclass of solutions of a Type II supergravity may feature a larger global symmetry than the whole set of solutions, namely than the theory itself. This is the case of solutions exhibiting $d$ coordinates which are associated with translational isometries, leaving the whole background invariant. At the classical level, this subclass is invariant under the action of an ${\rm O}(d,d)$ group, although no actual compactification is made along the $d$ directions. In fact, the invariance extends to ${\rm E}_{d+1(d+1)} $.\footnote{Recall that ${\rm E}_{3(3)}={\rm SL}(2,\mathbb{R})\times {\rm SL}(3,\mathbb{R})$,  ${\rm E}_{4(4)}={\rm SL}(5,\mathbb{R})$, and ${\rm E}_{5(5)}={\rm SO}(5,5)$. ${\rm E}_{d+1(d+1)}$ is the on-shell global symmetry of classical maximal ungauged supergravities in $(10-d)$-dimensions, once all form fields are dualised to lower-rank ones.} The group ${\rm O}(d,d)$ (or, the larger group ${\rm E}_{d+1(d+1)} $), can be used for generating new solutions from a given one in the same class. We shall restrict ourselves to the $d=2$ case \cite{Lunin:2005jy}. Given two cyclic coordinates $x_1,\,x_2$, a TsT transformation $(x_1,x_2;\gamma)$ is a ${\rm O}(2,2)$ transformation on the worldvolume theory, which results from the following, subsequent transformations on the pseudo vector $X_M=(\tilde{x}_i,\,x^i)$, $i=1,2$:
\begin{equation}
    x^1\,\stackrel{T}{\rightarrow}\,\,\tilde{x}_1\,\,,\,\,\,\, x^2\,\stackrel{s}{\rightarrow}\,\,x^2-\gamma\,\tilde{x}_1\,\,,\,\,\,\,  \tilde{x}_1\,\stackrel{T}{\rightarrow}\,\,\,x^{1\,\prime}\,.
\end{equation}
We split the ten-dimensional coordinates into $x^\mu=(x^\alpha,\,x^i)$ and suppose, for the sake of simplicity, that $B_{\alpha i}=G_{\alpha i}=0$. Then the transformation property of the components of the metric ${\bf G}\equiv (G_{ij})$ (in the string frame) and of the Kalb-Ramond field ${\bf B}\equiv (B_{ij}) $ is conveniently described by defining ${\bf E}\equiv {\bf G}+{\bf B}$, as follows \cite{Giveon:1994fu}:
\begin{equation}
    {\bf E}\rightarrow {\bf E}'={\bf E}\cdot (\boldsymbol{\gamma}\cdot {\bf E}+{\bf 1})^{-1}\,\,,\,\,\,\,\,\,\boldsymbol{\gamma}\equiv \left(\begin{matrix}0 & \gamma\cr -\gamma & 0\end{matrix}\right)\,.
\end{equation}
The dilaton transforms as: $e^{2\Phi'}=e^{2\Phi}/{\rm det}(\boldsymbol{\gamma}\cdot {\bf E}+{\bf 1})$. The transformation of the R-R fields is effected by grouping them in a polyform and acting on it by the corresponding ${\rm Spin}(2,2)$ transformation (see Appendix B of \cite{Maurelli:2025ueo} for a review of TsT transformations). The TsT transformation is in general a $\boldsymbol{\beta}$-transformation in the group ${\rm O}(d,d)$ \cite{Frolov:2005dj,Catal-Ozer:2022izq}, with $\boldsymbol{\beta}\equiv \boldsymbol{\gamma}$.
\section{The 2-Parameter Interpolating Solution}
We review here the construction of the 2-parameter, supersymmetric solution, {presented} in \cite{Maurelli:2025ueo}, which interpolates between a ${\rm WAdS}_3\times {\rm WS}^3\times T^4$ near-horizon geometry and {asymptotically-locally}-flat spacetime. For Type IIB supergravity, we follow the notation of \cite{Maurelli:2025ueo}, in which the NS-NS and R-R fields are denoted as follows:
\begin{align}
   {\rm NS-NS}:\,\,\, \left[g_{\mu\nu},\,\mathsf{B}_2=(B_{\mu\nu}),\,\Phi\right]\,\,;\,\,\,\, {\rm R-R}:\,\,\,\left[\mathsf{C}_0,\,\mathsf{C}_2=(C_{\mu\nu}),\,\mathsf{C}_4=(C_{\mu\nu\rho\sigma})\right]\, .\nonumber
\end{align}
The corresponding field strengths are defined in the following way:
\begin{align}
\mathsf{H}_{3}&=\dd \mathsf{B}_2 \,,\hspace{1.2cm}\mathsf{F}_{1}  =\dd \mathsf{C}_{0}\,,\hspace{1.2cm}\mathsf{F}_{3}=\dd \mathsf{C}_{2}-\mathsf{C}_{0}\wedge \mathsf{H}_{3}\,,\hspace{1.2cm}\mathsf{F}_{5}=\dd \mathsf{C}_{4}-\mathsf{C}_{2}\wedge \mathsf{H}_{3}={}^*\,\mathsf{F}_{5}\nonumber\,.
\end{align}
We refer to \cite{Maurelli:2025ueo} for the action and the supersymmetry transformations.\par
The starting point is the well-known D1-D5 system, consisting of stacks of parallel D1 and D5 branes, in which the latter, along the orthogonal directions to the 1-branes, are wrapped on $T^4$. We denote by $x_\pm$ the lightcone coordinates along the common directions of the D1 and D5 branes (the boundary of ${\rm AdS}_3$ in the near-horizon limit). The $\mathbb{R}^4$ space orthogonal to both branes is described by a radial coordinate $r$ and three angles $\psi,\varphi_1,\,\varphi_2$ spanning a 3-sphere $S^3$. Finally, the $T^4$ coordinates are denoted by $\vec{y}=(y^a)_{a=1,\dots, 4}$.
The solution, in the string frame and fixing the dilaton to vanish at infinity, reads
\begin{align}\
\dd s^{2}_{\mbox{\tiny s}} & =\sqrt{H_{1}H_{5}}\left(\dd r^{2}+\frac{r^{2}}{{4}}{[\dd\psi^{2}+\sin^{2}\psi\dd\varphi_{1}^{2}+(\dd\varphi_{2}-\cos\psi\dd\varphi_{1})^{2}]}\right)-\frac{\dd x_{+}\dd x_{-}}{\sqrt{H_{1}H_{5}}}+\sqrt{\frac{H_{1}}{H_{5}}}\,{|\dd\vec{y}|^2}\,,\nonumber\\
\mathsf{F}_{3} & =\frac{1}{8}d_{5}\sin\psi\dd\varphi_{1}\wedge\dd\varphi_{2}\wedge\dd\psi+\frac{d_{1}}{2r^{3}H_{1}^{2}}\dd r\wedge\dd x_{-}\wedge\dd x_{+}\,,\,\,\,\Phi =\frac{1}{2}\log\frac{H_{1}}{H_{5}}\,, \nonumber\\
\mathsf{H}_{3}  &=0\,,\,\,\mathsf{F}_{1}=0\,,\,\,\mathsf{F}_{5}=0\,,\,\,\,\,H_{1} =1+\frac{d_{1}}{2r^{2}}\,,\,\,\,H_{5} =1+\frac{d_{5}}{2r^{2}}\,.\label{D1D5}
\end{align}
The parameters, $d_1,\,d_5$ are proportional to the number of the two kinds of branes, and we take $d_1 d_5>0$. The solution is $1/4$-BPS. The near-horizon limit is effected by  rescaling $r\rightarrow \lambda r\,,\,\,x_\pm \rightarrow \sqrt{d_1 d_5}\,x_\pm/\lambda$ and taking the limit $\lambda\rightarrow 0$, which is isotropic in the boundary directions $x_\pm$. The resulting geometry is ${\rm AdS}_3\times S^3\times T^4$ and is $1/2$-BPS.
Next, we perform on \eqref{D1D5} two TsT transformations: $(x_+,y_4;\,\gamma_1)$ and $(\varphi_2,y_3;\,\gamma_2)$. The resulting solution reads:
\begin{align}
\dd s^{2} & =\sqrt{H_{1}H_{5}}\left[\dd r^{2}+\frac{r^{2}}{4}\left(\dd\psi^{2}+\sin^{2}\psi\,\dd\varphi_{1}^{2}+\frac{H_{5}}{H_{1}}e^{2\Phi}\mathbf{A}_{2}^{2}\right)\right]-\sqrt{\frac{H_{5}}{H_{1}}}\left(\mathbf{A}_{1}\dd x_{+}+\frac{\gamma_{1}^{2}}{4}\mathbf{A}_{1}^{2}\right) +\nonumber\\ &\,\,\,\,+\sqrt{\frac{H_{1}}{H_{5}}}\left(\dd y_{1}^{2}+\dd y_{2}^{2}+\frac{H_{5}}{H_{1}}e^{2\Phi}\dd y_{3}^{2}+\dd y_{4}^{2}\right)\,, \nonumber\\
\mathsf B_{2} & =\frac{\gamma_{1}}{2}\mathbf{A}_{1}\wedge\dd y_{4}-\frac{H_{5}\gamma_{2}r^{2}}{4}e^{2\Phi}\mathbf{A}_{2}\wedge\dd y_{3}\,,\qquad e^{2\Phi}=\frac{4H_{1}}{(4+r^{2}\gamma_{2}^{2}H_{1})H_{5}}\,, \nonumber\\
\mathsf{F}_{3} & =\frac{1}{8}(d_{5}\sin\psi\dd\varphi_{1}\wedge\dd\varphi_{2}\wedge\dd\psi+\frac{4d_{1}}{r^{3}H_{1}^{2}}\dd r\wedge\dd x_{-}\wedge\dd x_{+})\,,\qquad\mathsf{F}_{1}=0\,, \nonumber\\
\mathsf{F}_{5} & =\frac{\gamma_{1}d_{5}}{2}\mathbf{A}_{1}\wedge\left(\frac{1}{8}\sin\psi\mathbf{A}_{2}\wedge\dd y_{4}\wedge\dd\varphi_{1}\wedge\dd\psi+\frac{1}{r^{3}H_{5}}\dd r\wedge\dd y_{1}\wedge\dd y_{2}\wedge\dd y_{3}\right) \nonumber\\
& +\frac{\gamma_{2}d_{1}}{8}\left(-\sin\psi\dd y_{1}\wedge\dd y_{2}\wedge\dd y_{4}\wedge\dd\varphi_{1}\wedge\dd\psi+\frac{H_{5}^{2}}{rH_{1}^{2}}e^{2\Phi}\mathbf{A}_{1}\wedge\mathbf{A}_{2}\wedge\dd r\wedge\dd x_{+}\wedge\dd y_{3}\right)\,,\label{firstTsT}
\end{align}
where $\mathbf{A}_{1}  \equiv\frac{\dd x_{-}}{H_{5}}\,,\,\,\,\mathbf{A}_{2}  \equiv\dd\varphi_{2}-\cos\psi\dd\varphi_{1}$ will reduce to the two contact 1-forms on ${\rm WAdS}_3 $ and ${\rm WS}^3$, in the near-horizon limit, respectively. The warping of the first factor is described by the on/off parameter $\gamma_1$. The solution is $1/8$-BPS. It is asymptotically, locally flat and, for large $r$,  the string coupling is arbitrarily small: $g_s=e^{\Phi}\approx 2/(\gamma_2 r)+ O(1/r^3)$. The solution near radial infinity can thus be described within perturbative superstring theory. The brane content of this configuration is determined by computing the conserved charges from the fluxes. It consists of D1, D5 and D3 branes, with charges, respectively, proportional to:
\begin{align}
 &   \frac{1}{2\pi^{2}}\int_{S^{3}}\mathsf{F}_{3}  =d_{5}\,,\,\,\,\,\,\qquad
    \frac{1}{(2\pi)^{4}\,2\pi{}^{2}}\int_{S^{3}\times T^{4}}(\mathsf{F}_{7} -\mathsf B_2\wedge \dd C_4)=d_{1}\,,\nonumber\\
   & \frac{1}{(2\pi)^3 4\pi }\int_{(S^1)^3\times S^2}(\mathsf F_5-\mathsf B_2\wedge \mathsf F_3) = \frac{d_1 \gamma_2}{8}\,.
\end{align}
The $\mathsf{H}_3$ flux, on the other hand, describes an NS5-brane charge continuously distributed along the radial direction.
\par
The near-horizon limit is effected by taking a non-isotropic (i.e. non-relativistic along the boundary directions) rescaling
$ r\to \lambda r\,, \,\,\,x_+\to  x_+\,, \,\,\, x_-\to \frac{1}{\lambda^2} \,\,x_-$, followed by the limit $\lambda \to 0$.
The resulting background is ${\rm WAdS}^0_3\times {\rm WS}^3\times T^4$. Unlike the D1-D5 brane system, the near-horizon limit does not feature a supersymmetry enhancement, meaning that the resulting solution is still $1/8$-BPS.\par Let us also mention that the spacetime described by the interpolating solution \eqref{firstTsT} is a Kundt spacetime with a null, geodesic Killing vector ${\bf l}=2\partial/\partial x_+$.\footnote{See \cite{Chow:2009vt} for a study of Kundt solutions in TMG.}\par The interpolating solution discussed above can be derived within a $D=6$ $\mathcal{N}=(1,1)$ theory, describing the closed string sector of Type IIB superstring compactified on a $T^4/\mathbb{Z}_2$ orientifold and discussed in detail in \cite{Maurelli:2025iba}.
The bosonic sector of this six-dimensional truncation consists of: 16 scalar fields, coming from the metric moduli of $T^4$ and the internal components of  $\mathsf{C}
_2$ and parametrizing the space ${\rm SO}(4,4)/{\rm S}[{\rm O}(4)\times {\rm O}(4)]$; the dilaton; one 2-form originating from $\mathsf{C}
_2$; eight vector fields from the components of $\mathsf{B}_2$ and $\mathsf{C}_4$ with one leg and three legs along $T^4$, respectively.

Using the ${\rm O}(4)$ global symmetry of this six-dimensional truncation, the above, warped near-horizon geometry can be generalized and shown to be part of the following family of solutions:
\begin{align}
\dd s^{2} & =c_0\left(\frac{\dd {r}^{2}}{r^2}-r^2 \dd x_-\dd x_{+}+{\frac{1}{4}[\dd\psi^{2}+\sin^{2}\psi\dd\varphi_{1}^{2}]}\right)- c_1\mathsf{A}_{1}^2+{c_2\mathsf{A}_2^2}+{|\dd \vec{y}|^2}\,,\nonumber\\
\mathsf{H}_3&=k_1\,\dd \mathsf{A}_1\wedge \vec{W}\cdot \dd\vec{y}+k_2\,\dd \mathsf{A}_2\wedge \vec{X}\cdot \dd\vec{y}\,, \qquad \Phi=p_0\,,\nonumber\\
\mathsf{F}_3&=p_1 r\, \dd r\wedge \dd x_-\wedge \dd x_+-\frac{p_2}{4}\,\sin\psi\,\dd\psi\wedge \dd\varphi_1\wedge \dd\varphi_2\,,\nonumber\\
\mathsf{F}_5&=G_5 + \star G_5\,,\nonumber\\
G_5&=r\, \dd r\wedge \dd x_-\wedge \dd x_+\wedge \mathsf{A}_2\wedge \vec{X}\cdot \dd\vec{y}-\frac{1}{4}\,\sin\psi\dd\psi\wedge \dd\varphi_1\wedge \dd\varphi_2\wedge \mathsf{A}_1\wedge \vec{W}\cdot \dd\vec{y}\,,\nonumber \\
\star G_5 &= -\frac{1}{2 \sqrt{c_0 c_2}} (\dd \mathsf{A}_1 W^a + \dd \mathsf{A}_2 X^a) \frac{1}{3!} \epsilon_{a b c d} \wedge \dd y^{b}\wedge \dd y^{c}\wedge \dd y^{d}\,.
\end{align}
where $\mathsf{A}_1\equiv r^2\,\dd x_-$, $\mathsf{A}_2\equiv \dd\varphi_{2}-\cos\psi\dd\varphi_{1}$  and $\vec{X},\,\vec{W}$ are two 4-dimensional vectors. The field equations and the Bianchi identities restrict these solutions to two subclasses, preserving different amounts of supersymmetry, and defined by the following conditions on the parameters:
\begin{align}
   & \mbox{$1/8$-BPS:}\nonumber\\
    &\,k_1=-k_2\,,\,\,p_1=-p_2=\frac{1}{k_2}\,,\,\, p_0=\frac{1}{2}\log(4 c_0 c_2 k_2^2)\,,\,\, |\vec{W}|^2=\frac{c_1}{k_1^2}\,,\,\,|\vec{X}|^2=\frac{c_0-4c_2}{4k_2^2}\,,\nonumber\\
    &  \mbox{$1/4$-BPS:}\nonumber\\
    &  \vec{X}\cdot \vec{W}=0\,,\,\,  k_1=k_2\,,\,\, p_1= p_2=\frac{1}{k_2}\,,\,\,p_0=\frac{1}{2}\log(4 c_0 c_2 k_2^2)\,,\,\,|\vec{W}|^2=\frac{c_1}{k_2^2}\,,\,\, |\vec{X}|^2=\frac{c_0-4c_2}{4k_2^2}\,,\nonumber
    \end{align}
The near-horizon geometry of the interpolating solutions falls in the first branch, for $\vec{X}\cdot \vec{W}=0$.

\section{A New Class of Warped Solutions}\label{NCWS}
In this section, we present new (supersymmetric) solutions of the general form ${\rm WAdS}_3\times {\rm WS}^3\times T^4$, which are not described within the  $D=6$ $\mathcal{N}=(1,1)$ truncation, but rather in the maximal $\mathcal{N}=(2,2)$ model. These backgrounds feature all three kinds of warpings of the anti-de Sitter factor. Before giving their explicit form, let us highlight their main properties.
\begin{itemize}
\item The lightlike solution ${\rm WAdS}^0_3\times {\rm WS}^3\times T^4$ has three parameters $\alpha,\,\Omega$ and $p$, the first being the on/off parameter which describes the lightlike warping. These backgrounds are $1/8$-BPS as they preserve 4 of the 32 supercharges of the maximal $D=10$ theory, for generic values of the parameters.
\item  The spacelike solution ${\rm WAdS}^+_3\times {\rm WS}^3\times T^4$ depends on three parameters $\omega,\Omega,p$ and describes the spacelike analog of the solutions found in \cite{Hoare:2022asa}, extended to two independent warping parameters ($\omega,\Omega$). The solutions are 1/4-BPS for generic values of the parameters.
\item The timelike solution ${\rm WAdS}^-_3\times {\rm WS}^3\times T^4$ generalizes those found in \cite{Hoare:2022asa} to independent warpings of the two factors. It depends on three parameters $\omega,\,\Omega,\,p$, the first two describing the warpings of the anti-de Sitter and the sphere factors, respectively. These solutions are $1/4$-BPS and seem to provide a counterexample to the statement that backgrounds of this kind, with independent warping parameters, are non-supersymmetric. It is important, however, to point out that the solutions discussed in this section were not obtained through TsT transformations and the above statement is not contradicted if referred to backgrounds obtained through this kind of transformations.  \end{itemize}
The supersymmetry of the solutions crucially depends on the relative `chirality' of the two contact 1-forms $\mathsf{A}_1,\,\mathsf{A}_2$ in their respective three-dimensional spaces. This chirality is chosen so that, in the spacelike and timelike cases, for equal warping parameters $\Omega=\omega$, the following self-duality conditions on the six-dimensional space ${\rm WAdS}_3\times{\rm WS}^3$  hold
$$\star_6 \left(\Vol(\mathrm{WAdS}_{3})+\Vol(\mathrm{WS}^{3})\right)=\left(\Vol(\mathrm{WAdS}_{3})+\Vol(\mathrm{WS}^{3})\right)\,\,,\,\,\,\,\star_6 \dd \left(\mathsf{A}_1\wedge \mathsf{A}_2\right)=\dd \left(\mathsf{A}_1\wedge \mathsf{A}_2\right)\,.$$
Let us introduce here, the building blocks necessary to describe these geometries, that is, the dreibein $(e^a)$ for lightlike, spacelike and timelike $\mathrm{WAdS}_3$:
\begin{align}\label{vielbeins}
 \mbox{WAdS}^0&:\,\,\,\,\,   e^{0}  =\dd x_+-(1+4 \alpha)\frac{ \dd x_-}{4 u^2}\,,\qquad e^{1}=\frac{\dd u }{u}\,,\qquad e^{2}=\dd x_++(1-4 \alpha )\frac{ \dd x_-}{4 u^2}\,,\nonumber\\
  \mbox{WAdS}^+&:\,\,\,\,\,      e^{0}  =\frac{e^\frac{\omega}{4}}{2} \dd t \sinh \rho\,,\qquad e^{1}=\frac{e^{\frac{\omega}{4}}}{2}\dd\rho\,,\qquad e^{2}=\frac{e^{-\frac{3\omega}{4}}}{2}(\dd\theta-\cosh\rho\dd t)\,,\nonumber\\
   \mbox{WAdS}^-&:\,\,\,\,\,     e^{0} =\frac{1}{2}e^{-\frac{3\omega}{4}}(\dd t-\cosh\rho\dd v)\,,\qquad e^{1}=\frac{1}{2}e^{\frac{\omega}{4}}\dd\rho\,, \qquad e^{2}=\frac{1}{2}e^{\frac{\omega}{4}}\sinh\rho\dd v\,,
\end{align}
where the metric at the origin is ${\rm diag}(-1,1,1)$ and, in the lightlike case, the radial variable $u$ is different from the coordinate $r$ used in the previous section.
The dreibein for the warped sphere is
\begin{align}
\ell^{1} & =\frac{1}{2}e^{\frac{\Omega}{4}}\dd\psi\,,\qquad\ell^{2}=\frac{1}{2}e^{\frac{\Omega}{4}}\sin\psi\dd\varphi_{1}\,, \qquad\ell^{3}=\frac{1}{2}e^{-\frac{3\Omega}{4}}(\dd\varphi_{2}-\cos\psi\dd\varphi_{1})\,.
\end{align}
Moreover, in the following, we will denote the volume element of a space by $\Vol$, obtained as the wedge product of the corresponding vielbein.
\par

\subsection{Lightlike warping} \label{sect_lightlike}
The 10-dimensional metric, in the string frame, has the form:
\begin{align}
 \dd s_{10}^{2} & = \dd s^{2}(\mathrm{WAdS}_{3}^{0})+e^{-\frac{\Omega}{2}}\dd s^2(\mathrm{WS}^3) + e^{\Phi}\dd y^{a}\dd y^{a}\,,
\end{align}
where the line elements for $\mathrm{WAdS}_{3}^{0}$ and $\mathrm{WS}^3$ spaces are constructed in terms of the vielbein in \eqref{vielbeins} and with orientation $\epsilon_{x_+,u,x_-,\psi\varphi_{1}\varphi_{2}y_{1}y_{2}y_{3}y_{4}}=1$. The matter fields are given by
\begin{align}
\Phi & =\log(\phi_{0})\,,\nonumber\\
\mathsf{F}_{3} & ={\frac{2}{{\phi_{0}}}(}e^{-\Omega}\Vol(\mathrm{WAdS}_{3}^{0})+e^{-\frac{3\Omega}{4}}\Vol(\mathrm{WS}^{3}))+\sqrt{\frac{\alpha(1-e^{-2\Omega})}{\phi_0}}\dd\mathsf{A}_{1}\wedge\dd y^{4}\,, \nonumber\\
\mathsf{H}_{3} & =\frac p2\sqrt{\alpha}\dd(\mathsf{A}_{1}\wedge\mathsf{A}_{2})-\frac12\sqrt{\phi_0(e^{2\Omega}-1)}\dd\mathsf{A}_{2}\wedge w\,,\nonumber \\
G_{5} & =-\frac{e^{-\Omega}\sqrt{\alpha}}{2}\Delta\dd(\mathsf{A}_{1}\wedge\mathsf{A}_{2})\wedge\dd y^{1}\wedge\dd y^{2}+e^{-\Omega}\sqrt{\frac{(e^{2\Omega}-1)}{\phi_{0}}}\mathsf{A}_{2}\wedge\mathrm{Vol}(\mathrm{WAdS}_{3}^{0})\wedge w
\end{align}
where $\Delta=\sqrt{1-e^{2\Omega}p^2}$, $w=e^{-\Omega}\Delta\dd y^{3}-p\dd y^{4}$ and
\begin{align}
    \mathsf A_1=\frac{\dd x_-}{u^2}\,,\qquad \mathsf A_2=\dd\varphi_2-\cos\psi \dd \varphi_1\,.
\end{align}
The solution is real
provided that the parameters take values on the following sets
\begin{align}
|p|\leq e^{-\Omega}\,,\qquad\phi_{0}>0\,,\qquad\Omega\geq0\,.
\end{align}
As mentioned above, the solution is 1/8-BPS as there are 4 Killing spinors. In the $\Omega\rightarrow 0$ limit, the sphere becomes round but the solution is still 1/8-BPS. If, on the other hand, we set $\alpha=0$, the solution is ${\rm AdS}_3\times {\rm WS}^3\times T^4$ and becomes 1/4-BPS. If $\Omega=\alpha=0$ we recover the undeformed, 1/2-BPS background.

\subsection{Spacelike Warping} \label{sect_spacelike}
The metric for the spacelike warping solution is given by
\begin{align}
\dd s^{2} & =e^{\frac{\Omega}{2}}\dd s^{2}(\mathrm{WAdS}_{3}^{+})+e^{\frac{\omega}{2}}\dd s^{2}(\mathrm{WS}^{3})+e^{\Phi}\dd y^{a}\dd y^{a}\,,
\end{align}
with orientation $\epsilon_{t\rho \theta\psi\varphi_{1}\varphi_{2}y_{1}y_{2}y_{3}y_{4}}=1$. The remaining type IIB fields read as follows
\begin{align}\label{solspace}
\Phi & =\log\phi_{0}\,,\\
\mathsf{F}_{3} & =\frac{2e^{\frac{\Omega-\omega}{2}}}{\phi_{0}}[e^{\frac{\omega}{4}}\Vol(\mathrm{WAdS}_{3}^{+})+e^{\frac{\Omega}{4}}\Vol(\mathrm{WS}^{3})]-e^{-\frac{(\omega+\Omega)}{4}}\sqrt{\frac{\sinh(\omega-\Omega)}{2\phi_{0}}}\dd\mathsf{A}_{2}\wedge\dd y^{4}\,,\nonumber \\
\mathsf{H}_{3} & =-\frac{p}{4}\dd(\mathsf{A}_{1}\wedge\mathsf{A}_{2})-e^{\frac{\omega+\Omega}{4}}\sqrt{\frac{\phi_{0}\sinh(\omega-\Omega)}{2}}\dd\mathsf{A}_{1}\wedge w_{1}\,,\nonumber \\
\mathsf{F}_{5} & =G_{5}+\star G_{5}\,,\nonumber \\
G_{5} & =\frac{1}{4}e^{-\frac{\omega+\Omega}{2}}\Delta\dd(\mathsf{A}_{1}\wedge\mathsf{A}_{2})\wedge\dd y^{1}\wedge\dd y^{2}+e^{-\frac{5}{4}\omega}\sqrt{\frac{2\sinh(\omega-\Omega)}{\phi_{0}}}\mathsf{A}_{1}\wedge\Vol(\mathrm{WS}^{3})\wedge w_{2}\,,\nonumber
\end{align}
where
\begin{align}
    \mathsf{A}_{1} & =\dd \theta-\cosh\rho\dd t\,,\hspace{3.3cm}\mathsf{A}_{2}=\sqrt{e^{2 \Omega }-1}(\dd\varphi_{2}-\cos\psi\dd\varphi_{1})\,,\\
    w_1&=e^{-\frac{\omega+\Omega}{2}}\Delta\dd y^{3}-pe^{\Omega}\dd y^{4}\,,\hspace{2cm}w_{2}=e^{\frac{\omega+\Omega}{2}}\Delta\dd y^{3}-pe^{\omega}\dd y^{4}\,,
\end{align}
and $\Delta=\sqrt{1-e^{\omega+\Omega}p^{2}}$. The solution is real provided that
\begin{align}
    \Omega\geq0\,,\qquad \omega\geq\Omega\,,\qquad |p|\leq e^{-\frac{\omega+\Omega}{2}}\,,\qquad \phi_0>0\,.
\end{align}
Being $\omega\ge 0$, the ${\rm WAdS}_3^+$ is of \emph{squashed} type.
Differently from the doubly deformed spacelike solution in \cite{Maurelli:2025iba}, the above background is not symmetric in the exchange of the respective deforming parameters, as it can be ascertained from the expression of the contact 1-form for the warped sphere. {This asymmetry manifests in a sort of hierarchy between the two deformation parameters: the limit $\omega\to0$ constraints $\Omega$ to be zero as well, while the round sphere limit can be taken for any value of $\omega$.}
In the limit $|p|=e^{-\frac{\omega+\Omega}{2}}$, the coefficient $\Delta$ vanishes, while further identifying the deforming parameters reduces the matter configuration to the one in \cite{Hoare:2022asa}, that is
\begin{align}
    \Phi & =\log\phi_{0}\,,\nonumber\\
    \mathsf{F}_{3} & =\frac{2e^{\frac{\omega}{4}}}{\phi_{0}}[\Vol(\mathrm{WAdS}_{3}^{+})+\Vol(\mathrm{WS}^{3})]\,,\nonumber \\
    \mathsf{H}_{3} & =-e^{-\omega}\dd(\mathsf{A}_{1}\wedge\mathsf{A}_{2})\,,\nonumber \\
    \mathsf{F}_{5} & =0\,.
\end{align}
{In this limit, both $\mathsf F_3$ and $\mathsf H_3$ become selfdual forms in six dimensions.}  The full solution \eqref{solspace} is 1/4 BPS, as it preserves eight supercharges, for any allowed limit of the parameters, except for the undeformed case in which the solution clearly becomes $1/2$-BPS.

\subsection{Timelike warping} \label{sect_timelike}
Inspired by the structure of \eqref{solspace}, one can construct a similar solution for the timelike case. Following again the notation in \eqref{vielbeins}, the total solution is given by
\begin{align}
\dd s^{2} & =e^{\frac{\Omega}{2}}\dd s^{2}(\mathrm{WAdS}_{3}^{-})+e^{\frac{\omega}{2}}\dd s^{2}(\mathrm{WS}^{3})+e^{\Phi}\dd y^{a}\dd y^{a}\,,\nonumber\\
\Phi & =\log\phi_{0}\,,\\
\mathsf{F}_{3} & =-\frac{2e^{\frac{\omega-\Omega}{2}}}{\phi_{0}}[e^{\frac{\omega}{4}}\Vol(\mathrm{WAdS}_{3}^{-})+e^{\frac{\Omega}{4}}\Vol(\mathrm{WS}^{3})]+e^{-\frac{(\omega+\Omega)}{4}}\sqrt{\frac{\sinh(\Omega-\omega)}{2\phi_{0}}}\dd\mathsf{A}_{1}\wedge\dd y^{4}\,,\nonumber \\
\mathsf{H}_{3} & =-\frac{p}{4}\dd(\mathsf{A}_{1}\wedge\mathsf{A}_{2})-e^{\frac{\omega+\Omega}{4}}\sqrt{\frac{\phi_{0}\sinh(\Omega-\omega)}{2}}\dd\mathsf{A}_{2}\wedge w_{1}\,,\nonumber \\
\mathsf{F}_{5} & =G_{5}+\star G_{5}\,,\nonumber \\
G_{5} & =\frac{1}{4}e^{-\frac{\omega+\Omega}{2}}\Delta\dd(\mathsf{A}_{1}\wedge\mathsf{A}_{2})\wedge\dd y^{1}\wedge\dd y^{2}-e^{-\frac{5}{4}\Omega}\sqrt{\frac{2\sinh(\Omega-\omega)}{\phi_{0}}}\mathsf{A}_{2}\wedge\Vol(\mathrm{WAdS}^{-}_{3})\wedge w_{2}\,,\nonumber
\end{align}
with orientation $\epsilon_{t\rho v\psi\varphi_{1}\varphi_{2}y_{1}y_{2}y_{3}y_{4}}=1$ and
\begin{align}
\mathsf{A}_{1} & =\sqrt{1-e^{2\omega}}(\dd t-\cosh\rho\dd v)\,,\hspace{1.5cm}\mathsf{A}_{2}=\dd\varphi_{2}-\cos\psi\dd\varphi_{1}\,,\\
w_{1} & =e^{-\frac{\omega+\Omega}{2}}\Delta\dd y^{3}+pe^{\omega}\dd y^{4}\,,\hspace{2cm}w_{2}=e^{\frac{\omega+\Omega}{2}}\Delta\dd y^{3}+pe^{\Omega}\dd y^{4}\,.
\end{align}
The parameter $\Delta$ is defined, as for the spacelike solution, as $\Delta=\sqrt{1-e^{\omega+\Omega}p^{2}}$. The total configuration is real when
\begin{align}
\Omega\geq\omega,\qquad\omega\leq0\,,\qquad|p|\leq e^{-\frac{\omega+\Omega}{2}}\,\qquad \phi_0>0\,.
\end{align}
This background is the generalization of those constructed in \cite{Hoare:2022asa} in that the warpings of the two factors are independent.
It reduces to the background of \cite{Hoare:2022asa} (modulo the action of the  ${\rm SO}(5)$-duality group used in the same reference) choosing $\Delta=0$, i.e. $p=e^{-\frac{\omega+\Omega}{2}}$, and $\Omega=\omega$.
As opposed to the spacelike case, the undeformed limits $\omega\to0$ and $\Omega\to 0$ can be taken independently, due to the negativity of $\mathrm{WAdS}_3$ parameter.
As for the solutions in \cite{Hoare:2022asa}, the above backgrounds are $1/4$-BPS. This holds for all choices of the parameters, except for the undeformed case in which the solution is 1/2-BPS. Following the discussion in that paper, one can construct similar solutions by exchanging appropriately $\Vol(\mathrm{WS^3})\leftrightarrow \dd(\mathsf{A}_1\wedge \mathsf{A_2})$.

\section{Conclusions}
We have presented here a selection of the results of \cite{Maurelli:2025ueo}, on the construction of backgrounds of the form ${\rm WAdS}_3\times {\rm WS}^3\times T^4$ and their realization as the near-horizon geometry of a configuration of branes and fluxes, which interpolates between this background and a locally flat spacetime within perturbative superstring theory. We have also presented new, supersymmetric solutions of this kind, with generic warpings of the anti-de Sitter factor. Important tools in the derivation of the former class of solutions, as well as warped-AdS spacetimes in general, are the TsT transformations. As mentioned earlier, these are part of a larger solution-generating (classical) symmetry group ${\rm E}_{3(3)}$, in the presence of two translational isometries. \par
An interesting project would be the construction of new backgrounds obtained using the \emph{hidden symmetries} in ${\rm SL}(3,\mathbb{R})\subset {\rm E}_{3(3)}$. These were already used in \cite{Lunin:2005jy} and introduce two parameters in the doublet representation of ${\rm SL}(2,\mathbb{R})_{\rm IIB}$. The interpretation of these deformations in the dual dipole-CFT is, to our knowledge, an open problem. On the other hand, an extension of this kind of the ${\rm O}(3,3)$-TsT transformations of \cite{Frolov:2005dj,Catal-Ozer:2022izq}, along three directions, is to be looked for among the hidden symmetries of ${\rm E}_{4(4)}\equiv {\rm SL}(5,\mathbb{R})$. All these issues, as well as the study of the global properties of the deformation parameters, are the subject of a current investigation \cite{inprogress}.\par
Another interesting problem is to study the stability of the non-supersymmetric doubly-warped backgrounds ${\rm WAdS}_3\times {\rm WS}^3\times T^4$, in which supersymmetry breaking is triggered by the continuous parameter describing the warping on any of the two factors \cite{inprogress2}. Such backgrounds include some of the 2-parameter ones found in \cite{Maurelli:2025iba} with spacelike and timelike warpings of the anti-de Sitter factor: they are generically non-BPS, but become supersymmetric as one of the two warping parameters is sent to zero.
It is indeed reasonable to expect that, within a small enough range of values of the supersymmetry-breaking parameter, perturbative stability of the solution should be preserved. Evidence in this direction was obtained by \cite{Eloy:2021fhc} in the case in which only the $S^3$ factor is continuously deformed to break all supersymmetries. Perturbatively stable, non-BPS ${\rm WAdS}_3\times {\rm WS}^3\times T^4$ backgrounds would represent variants of the (few) examples of potentially stable non-supersymmetric anti-de Sitter backgrounds and, as such, would have a role in the context of a possible extension of the AdS swampland conjecture \cite{Ooguri:2016pdq}  to WAdS backgrounds. The conjecture in \cite{Ooguri:2016pdq}, indeed, states that non-supersymmetric anti-de Sitter backgrounds are unstable.\par
Finally, it would be interesting to understand if the solutions discussed in Section \ref{NCWS} describe the near-horizon geometry of a suitable configuration of branes and fluxes in the Type IIB theory.
\vspace{0.3cm}

\noindent\textbf{Public code.} The solutions presented in the sections \ref{sect_lightlike}, \ref{sect_spacelike} and \ref{sect_timelike} can be verified using the Wolfram script available at

\hfill
\url{https://github.com/moyarzoca/typeIIB-warped-solutions-checks}.

\section{Acknowledgements}
We would like to thank Domenico Orlando, Ivo Sachs, Henning Samtleben, Alessandro Sfondrini, Ricardo Stuardo and Konstantinos Zoubos for fruitful discussions. R.N. was supported by the European Union and the Czech Ministry of Education, Youth and Sports
(Project: MSCA Fellowship CZ FZU III - CZ.02.01.01/00/22 010/0008598). MO is supported by AEI-Spain PID2023-152148NB-I00 and by Maria de Maeztu excellence unit grant CEX2023-001318-M, by Xunta de Galicia (CIGUS Network of Research Centres and project ED431F-2023/19), and by the European Union FEDER.

\bibliographystyle{JHEP}

\end{document}